# Focusing of Rayleigh waves generated by high-speed trains under the condition of ground vibration boom

Victor V. Krylov

Department of Aeronautical and Automotive Engineering,
Loughborough University,
Loughborough, Leicestershire  LE11 3TU, UK

**Abstract**

In the present paper, the effects of focusing of Rayleigh waves generated by high speed trains in the supporting ground under the condition of ground vibration boom are considered theoretically. These effects are similar to the effects of focusing of sound waves radiated by aircraft under the condition of sonic boom. In particular, if a railway track has a bend, e.g. to provide the possibility of changing direction of train movement, the Rayleigh surface waves generated by high-speed trains under the condition of ground vibration boom may become focused. This results in concentration of their energy along a simple caustic line at one side of the track and in the corresponding increase in ground vibration amplitudes. The effect of focusing of Rayleigh waves may occur also if a train moves along a straight line with acceleration $a$ and its current speed $v(t)$ is higher than Rayleigh wave velocity in the ground. In the present paper, both the above-mentioned focusing mechanisms are investigated in detail using the Green's function formalism and the expressions for space-time distributions of wheel-axle loads that take into account either track curvature or train acceleration. It is shown that in both these cases the effect of focusing can result in noticeable increase in generated ground vibrations. The obtained results are illustrated by numerical calculations.

## 1  Introduction

High-speed trains represent a convenient and environmentally friendly alternative to road and air transportation. The last two decades have been marked by rapid development of high-speed railway systems in many countries throughout the world. As many other means of transportation, high-speed trains are not free of environmental problems. In particular, ground vibrations generated by high-speed trains is one of the major environmental problems that must be mitigated to allow high-speed trains to be used in densely populated areas.

    It is well known that, if train speeds increase, the intensity of railway-generated vibrations generally becomes larger. For modern high-speed trains the increase in generated ground vibrations is especially high when train speeds approach certain critical velocities of elastic waves propagating in a track-ground system, the most important of them being the velocity of Rayleigh surface wave in the supporting ground. As has been theoretically predicted by the

present author [1- 4], if a train speed $v$ exceeds the Rayleigh wave velocity $c_R$ in supporting soil a *ground vibration boom* occurs. This phenomenon is similar to a sonic boom for aircraft crossing the sound barrier, and it is associated with a very large increase in generated ground vibrations, as compared to the case of conventional trains. The existence of ground vibration boom has been later confirmed experimentally [5, 6], which implies that one can speak of 'supersonic' ('superseismic') or, more precisely, 'trans-Rayleigh' trains [7-9] in the same way as people speak of supersonic aircraft. The increased attention to the problems of ground vibrations associated with high-speed trains is reflected in a growing number of theoretical and experimental investigations in this area (see, e.g. [10-18]).

The aim of the present paper is to investigate the effect of focusing of ground vibrations generated by high-speed trains that may take place under the condition of ground vibration boom, i.e. at trans-Rayleigh train speeds, even in a homogeneous ground. Some preliminary results on this topic have been reported earlier [19]. The two specific cases to be considered are focusing due to track curvature and focusing due to train acceleration, which are of the same physical nature as similar cases of focusing of a sonic boom from supersonic aircraft, sometimes called 'superbooms' [20-22]. It will be demonstrated that in both these cases the effect of focusing can result in noticeable increase in generated ground vibrations.

## 2  Outline of the theory of railway-generated ground vibrations

The main mechanisms of railway-generated ground vibrations are the wheel-axle pressure onto the track, the effects of joints in unwelded rails, and the dynamically induced forces of carriage- and wheel-axle vibrations excited due to unevenness of wheels and rails. In this paper we consider only the most common generation mechanism that is present even for ideally flat rails and wheels – namely, the quasi-static pressure of wheel axles onto the track, which is also responsible for railway-generated ground vibration boom.

According to the earlier developed general theory [1-4, 7-9], in order to calculate ground vibrations generated by a train due to the quasi-static pressure mechanism, one needs to take into account the superposition of waves generated by each elementary source of ground vibrations (sleeper) activated by wheel axles of all carriages, with the time and space differences between sources (sleepers) being taken into account. Using the Green's function of an elastic half space $G_{zz}(\rho,\omega)$, the frequency spectrum of the normal component of ground vibration velocity on the ground surface $v_z(x, y, \omega)$ can be written as

$$v_z(x,y,\omega) = \int_{-\infty}^{\infty}\int_{-\infty}^{\infty} P(x',y',\omega) G_{zz}(\rho,\omega) dx' dy', \qquad (1)$$

where $P(x',y',\omega)$ is the Fourier spectrum of distributed dynamic forces acting from all sleepers to the ground, and $\rho$ is the distance from each sleeper to the point of observation characterised by the coordinates $x, y$. The expression for the Green's function in (1), in which we take into account only the contribution of generated Rayleigh waves, can be written in the form (see e.g. [9, 23, 24])

$$G_{zz}(\rho,\omega) = D(\omega)\frac{1}{\sqrt{\rho}}\exp(ik_R\rho - \gamma k_R\rho), \qquad (2)$$

where

$$D(\omega) = \frac{(-i\omega)q k_R^{1/2} k_t^2}{(2\pi)^{1/2} \mu F'(k_R)} \exp(-i\frac{3\pi}{4}),  \quad (3)$$

and the factor $F'(k_R)$ is a derivative of the so-called Rayleigh determinant

$$F(k) = (2k^2 - k_t^2)^2 - 4k^2(k^2 - k_t^2)^{1/2}(k^2 - k_l^2)^{1/2} \quad (4)$$

taken at $k = k_R$. The structure of the function $P(x',y',\omega)$, i.e. the Fourier transform of the spatial distribution of time-dependent dynamic forces acting from sleepers to the ground, will be discussed below.

The notations in (1)-(4) are as follows: $\rho = [(x-x')^2 + (y-y')^2]^{1/2}$ is the distance between the source (with current coordinates $x', y'$) and the point of observation located on the surface (with the coordinates $x, y$), $\omega$ is a circular frequency, $k_R = \omega/c_R$ is the wavenumber of a Rayleigh surface wave, $c_R$ is the Rayleigh wave velocity, $k_l = \omega/c_l$ and $k_t = \omega/c_t$ are the wavenumbers of longitudinal and shear bulk elastic waves, where $c_l = [(\lambda + 2\mu)/\rho_0]^{1/2}$ and $c_t = (\mu/\rho_0)^{1/2}$ are longitudinal and shear wave velocities, $\lambda$ and $\mu$ are the elastic Lamé constants, $\rho_0$ is mass density of the ground, $q = (k_R^2 - k_l^2)^{1/2}$, and $\gamma = 0.001 - 0.1$ is a non-dimensional loss factor describing the attenuation of Rayleigh waves in soil.

As was mentioned above, function $P(x',y',\omega)$ describes the frequency spectrum of the spatial distribution of all load forces acting along the track. This spectrum can be found by taking a Fourier transform of the time and space dependent load forces $P(t, x', y')$ applied from the track to the ground. Note that the function $P(t, x', y')$ does not depend on possible layered structure of the ground and remains the same for both vertically homogeneous and inhomogeneous half spaces. In the model under consideration, all properties of track and train, which determine generation of ground vibrations, are described by the above mentioned function of load forces $P(t, x', y')$.

Being interested in fundamental features of the phenomenon of focusing of railway-generated Rayleigh waves, in this paper, for the sake of simplicity, we consider ground vibrations generated by a single axle load only. We recall that for a single axle load moving at speed $v$ along a straight track (located at $y = 0$), the load function has the form [2, 9]:

$$P(t, x', y'=0) = \sum_{m=-\infty}^{\infty} P(t - x'/v)\delta(x'-md)\delta(y'), \quad (5)$$

where $P(t-x'/v)$ is the time-delayed dynamic force acting from a sleeper with a coordinate $x'$ to the ground surface, and the delta-function $\delta(x'-md)$ takes the periodic distribution of sleepers into account. Using the expression for $P(t)$ (see [4, 8, 9]), taking the Fourier transform of (5) to calculate $P(x',y',\omega)$, and substituting the result into (1) using the expression for Green's function $G_{zz}(\rho,\omega)$ (see formulas (2)-(4)) results in the following expression for the vertical vibration velocity $v_z$ of Rayleigh waves generated on the ground surface ($z = 0$) at the point of observation with the coordinates $x, y$ by a single axle load moving along the straight track at speed $v$:

$$v_z(x, y, \omega) = P(\omega)D(\omega) \sum_{m=-\infty}^{\infty} \frac{1}{\sqrt{\rho_m}} \exp\left[i\frac{\omega}{v}md + (i-\gamma)\frac{\omega}{c_R}\rho_m\right]. \quad (6)$$

Here $\rho_m = [(x - md)^2 + y^2]^{1/2}$ is the distance from the sleeper characterised by the number $m$ to the observation point. The function $P(\omega)$ in (6) has the following form (see [4, 8, 9] for detail):

$$P(\omega) = -\frac{12.8\frac{Td}{v\pi^2}}{\frac{\omega^4}{\beta^4 v^4} - 4\frac{\omega^2}{c_{min}^2 \beta^2} - 8i\frac{g\omega}{c_{min}\beta} + 4}, \quad (7)$$

where $T$ is the axle load, $c_{min}$ is the minimal phase velocity of track flexural waves propagating in a track/ground system, $\beta$ is the parameter dependent on the elastic properties of track and ground and measured in $m^{-1}$, and $g$ is a non-dimensional track damping parameter. For relatively low train speeds, i.e., for $v < c_R$, the dynamic solution (7) for the force spectrum $P(\omega)$ goes over to the quasi-static one [8, 9]. As train speeds increase and approach or exceed the minimal track wave velocity, the spectra $P(\omega)$ become broader and larger in amplitudes, and a second peak appears at higher frequencies.

For 'trans-Rayleigh trains', i.e. for trains travelling at speeds $v$ higher than Rayleigh wave velocity in the ground $c_R$, the analysis of the expression (6) shows that maximum radiation of ground vibrations takes place if the train speed $v$ is larger than Rayleigh wave velocity $c_R$ [1-4, 7-9]. Under this condition, a *ground vibration boom* takes place, i.e., ground vibrations are generated as quasi-plane Rayleigh surface waves symmetrically propagating at angles $\Theta = cos^{-1}(c_R/v)$ in respect to the track, and with amplitudes much larger than in the case of conventional trains. Note that for trans-Rayleigh trains these Rayleigh surface waves are generated equally well on tracks with and without railway sleepers, whereas for conventional trains the presence of sleepers is paramount. Without them no propagating waves are generated in the framework of the quasi-static pressure generation mechanism.

## 3  Focusing due to track curvature

If a track has a bend of radius $R$ to provide the possibility of changing direction of train movement, the wave fronts of Rayleigh ground waves generated under the condition of ground vibration boom may become concave at one side of the track, instead of being convex (under usual circumstances) or plane (under the condition of ground vibration boom). This may result in focusing of generated ground vibrations (Rayleigh surface waves) along a simple caustic line at one side of the track accompanied by the corresponding increase in their amplitudes. According to the geometrical acoustics approximation, this increase is up to infinity. However, because of the diffraction limit, the real increase for this type of caustic is much more modest and does not exceed 2-4 times.

It can be easily shown that if a train moves along a circular bend of radius $R$ at speed $v$ (with $v > c_R$) then the caustic line formed by rays of Rayleigh waves radiated at Mach angles $\Theta$ is a concentric circle of a smaller radius $r$ (see Figure 1). Indeed, one can see from Figure 1 that $r = R\sin(90^0 - \theta) = R\cos\theta$. On the other hand, as it was mentioned in the previous section, under the condition of ground vibration boom $\cos\Theta = c_R/v$. Therefore, the radius $r$ of the concentric circle forming the caustic line is defined by the obvious simple expression:

$$r = R\frac{c_R}{v} \ . \tag{8}$$

The wave analysis of this problem is rather straightforward and is based on the rewriting the expression for a distributed axle loads that takes into account the two-dimensional geometry of a track with curvature. Namely, for a curved track, the expression (5) should be modified following the geometry of the curved track shown in Figure 2.

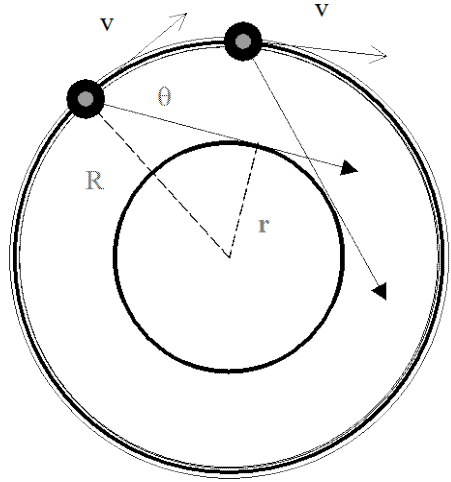

Figure 1. Geometrical acoustics explanation of the focusing of Rayleigh waves radiated by a train (a single axle load in this example) travelling at speed $v$ along a curved track of radius $R$ under the condition of ground vibration boom ($v > c_R$). The focusing occurs along the caustic line formed by a concentric circle of smaller radius $r$.

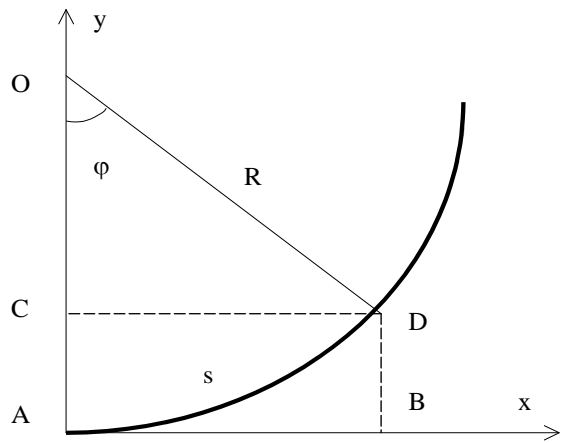

Figure 2. Geometry of the problem of the curved track.

As it follows from Figure 2, $AB = CD = R\sin\varphi$, $OC = R\cos\varphi$, and $DB = AC = OA - OC = R(1 - \cos\varphi)$. Therefore, instead of the above-mentioned formula (5) for $P(t, x', y' = 0)$, the following expression for $P(t, x', y')$ should be used:

$$P(t, x', y') = \sum_{m=-\infty}^{\infty} P(t - \frac{s'}{v})\delta(x' - R\sin\frac{md}{R})\delta(y' - R(1 - \cos\frac{md}{R})), \quad (9)$$

where the coordinate $s'$ is measured along the curved track. Taking the Fourier transform of (9), substituting it into (1) and taking into account that for a sleeper with the number $m$ the angle $\varphi = \varphi_m = md/R$, where $d$ is a sleeper periodicity, one can obtain the following formula for the spectral component of vertical vibration velocity $v_z(x,y,\omega)$ of Rayleigh waves generated on the ground surface at the point of observation with the coordinates $x, y$ by a single axle load moving along the curved track at speed $v$:

$$v_z(x, y, \omega) = P(\omega)D(\omega) \sum_{m=-\infty}^{\infty} \frac{1}{\sqrt{\rho^c_m}} \exp\left[i\frac{\omega}{v}md + (i - \gamma)\frac{\omega}{c_R}\rho^c_m\right]. \quad (10)$$

Here the distance $\rho^c_m$ from the sleeper characterised by the number $m$ to the observation point is described by the formula:

$$\rho^c_m = \sqrt{\left[x - R\sin\left(\frac{md}{R}\right)\right]^2 + \left[y - R(1 - \cos\left(\frac{md}{R}\right))\right]^2}, \quad (11)$$

and the other notations are the same as in the previous section.

The results of the calculations of the spatial distribution of ground vibration field $v_z(x,y,\omega)$ over the surface area of $75 \times 75$ m (in arbitrary units) generated at the frequency component $f = 10$ Hz by a single axle load travelling along the curved track with the radius of curvature $R = 100$ m are shown in Figure 3. The direction of travel is from left to right. The load speed is $v = 50$ m/s, and the velocity of Rayleigh wave in the ground was set as $c_R = 45$ m/s. Other relevant parameters were as follows: $d = 0.7$ m, $\beta = 1.28$ m$^{-1}$, $\gamma = 0.001$.

It can be clearly seen from Figure 3 that the wave fronts of generated Rayleigh waves become concave on the left-hand side of the track (in respect of the direction of travel), which results in focusing of Rayleigh waves along the caustic line accompanied by the increase in their amplitudes. Numerical values, that are not displayed here for clarity, show that the average increase in amplitudes due to focusing is about two times.

Figure 4 shows the calculated spatial distribution of ground vibrations $v_z(x, y, \omega)$ over the surface area of $75 \times 75$ m generated by a single axle load travelling along the curved track with the larger radius of curvature $R = 200$ m. All other parameters are the same as in Figure 3. Like in Figure 3, the location of the curved track can be easily identified in Figure 4. One can also see from Figure 4 that the focusing of generated Rayleigh waves, although less pronounced than in the case of $R = 100$ m, is still clearly visible on the left from the curved track relative to the direction of travel, and the caustic is located further away from the track, as expected from the expression (8).

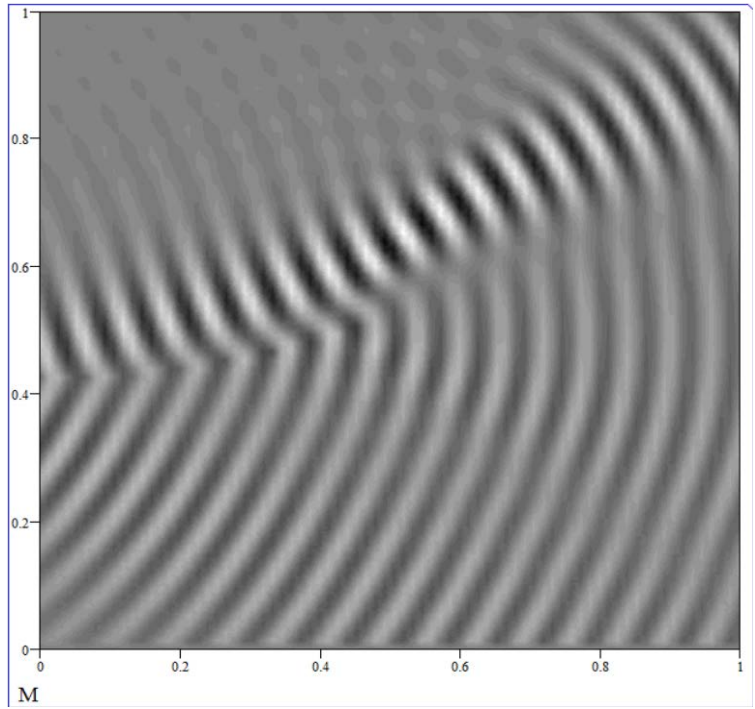

Figure 3. Spatial distribution of ground vibration field (in arbitrary units, greyscale) generated over the area of 75×75 m by a single axle load travelling at speed $v = 50\ m/s$ on a curved track with the radius of curvature $R = 100\ m$.

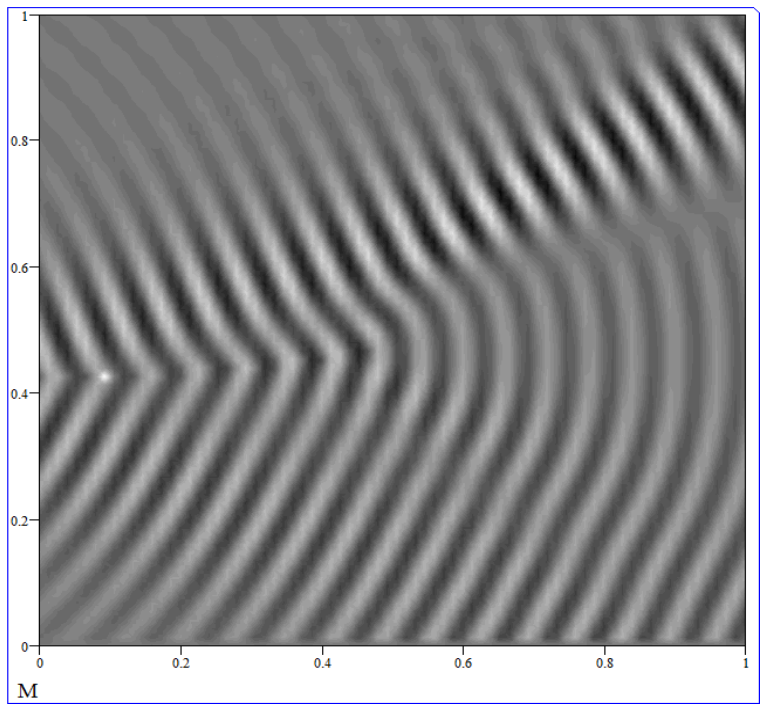

Figure 4. Spatial distribution of ground vibration field (in arbitrary units, greyscale) generated over the area of 75×75 m by a single axle load travelling at speed $v = 50\ m/s$ on a curved track with the radius of curvature $R = 200\ m$.

Note that the radii of track curvature *R = 100 m* and *R = 200 m*, for which the results of Figures 3 and 4 have been calculated, are rather small for real high-speed trains that require much larger radii of curvature to avoid derailment (e.g. for a train speed of 33 m/s (or 120 km/h) the value of *R* should be not less than 450 m). However, using the unrealistically small values of *R = 100 m* and *R = 200 m* for the example calculations presented in Figures 3 and 4 helps to illustrate the phenomenon of focusing due to track curvature more clearly.

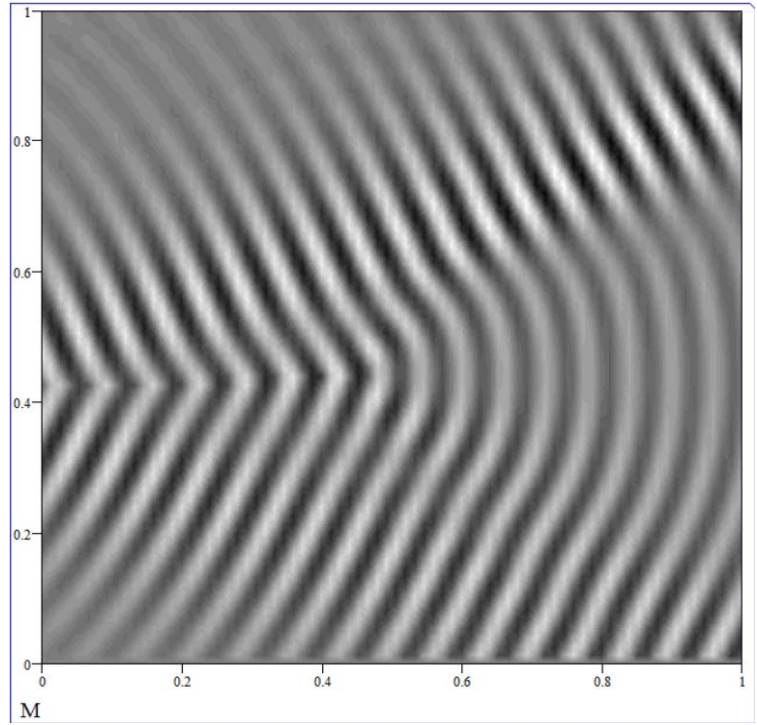

Figure 5. Spatial distribution of ground vibration field (in arbitrary units, greyscale) generated over the area of 75×75 m by a single axle load travelling at speed *v = 50 m/s* on a curved track with the radius of curvature *R = 400 m*.

Figure 5 shows the calculated spatial distribution of ground vibrations $v_z(x, y, \omega)$ over the surface area of $75 \times 75$ m generated by a single axle load travelling along the curved track with the larger and more realistic radius of curvature *R = 400 m*. All other parameters are the same as in Figures 3 and 4. One can see from Figure 5 that the formation of concave wave fronts and the focusing are still visible on the left from the curved track relative to the direction of travel (the load is moving from left to right). Although the associated increase in amplitudes of generated Rayleigh ground waves for Figure 5 (as well as for Figures 3 and 4) is only about two times (6 dB), it should be kept in mind that this two-fold increase is in addition to the already very large level of generated vibrations due to the ground vibration boom. Railway engineers and city planners should be aware of this phenomenon in order to avoid construction of residential buildings or placing sensitive equipment in the locations of the expected focusing of generated Rayleigh waves.

## 4 Focusing due to train acceleration

The effect of focusing of Rayleigh waves generated by high-speed trains may occur also in the case of a train accelerating along a straight line if its current speed, $v = v(t) = v_0 + at$, is higher than Rayleigh wave velocity in the ground $c_R$ (here $v_0$ is the initial train speed and $a$ is the acceleration). The geometrical acoustics consideration of this effect is illustrated in Figure 6 for three different positions of a moving train characterised by the changed angles of radiation of Rayleigh wave rays - angles $\theta_1$, $\theta_2$ and $\theta_3$ respectively. One can see that the caustic line in this case is not confined to the area near the train path, but moves away from it as the train moves along the track and its speed increases.

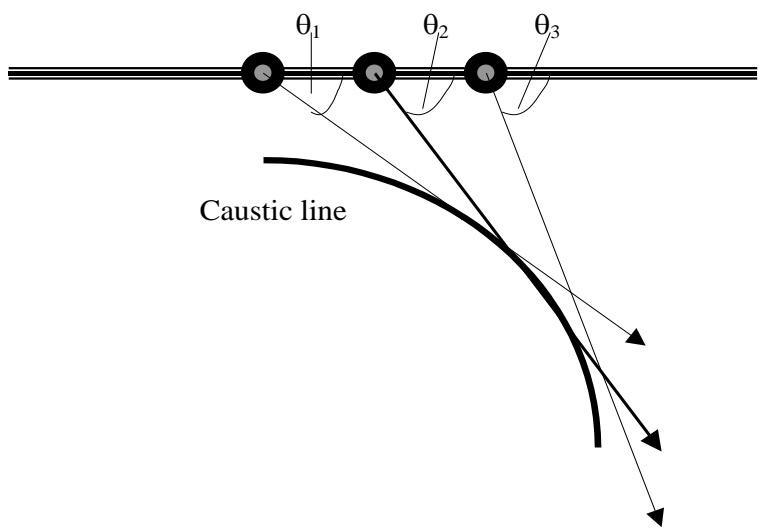

Figure 6. Geometrical acoustics explanation of the focusing of Rayleigh waves radiated by a train (a single axle load in this example) travelling with constant acceleration $a$ along a straight track under the condition of ground vibration boom ($v > c_R$). The focusing occurs along the caustic line that is moving away from the track as the train passes by with acceleration.

To apply the wave approach to this problem, it is convenient to express the train current speed $v$ as a function of the distance $s$ measured along the track. It follows from the kinematics of motion of a particle with acceleration $a$ that

$$v = v_0 \sqrt{1 + \frac{2as}{v_0^2}}. \qquad (12)$$

Expressing the distance as $s = md$ and substituting the expression (12) for $v$ into (6) gives the following formula for ground vibration velocity generated by a single load moving with acceleration $a$:

$$v_z(x,y,\omega) = P(\omega)D(\omega) \sum_{m=-\infty}^{\infty} \frac{1}{\sqrt{\rho_m}} \exp\left[ i \frac{\omega}{v_0\sqrt{1-2amd/v_0^2}} md + (i-\gamma)\frac{\omega}{c_R}\rho_m \right]. \qquad (13)$$

The meaning of the other functions and parameters in the expression (13) remain the same as in the previous sections.

The results of calculations of the spatial distribution of ground vibrations over the area of 75×75 m generated at the frequency component $f = 10\ Hz$ by a single axle load travelling along the straight track with a typical train acceleration $a = 0.5\ m/s^2$ are shown in Figure 7. The initial load speed was $v_0 = 50\ m/s$, and the velocity of Rayleigh wave in the ground was set as $c_R = 45\ m/s$.

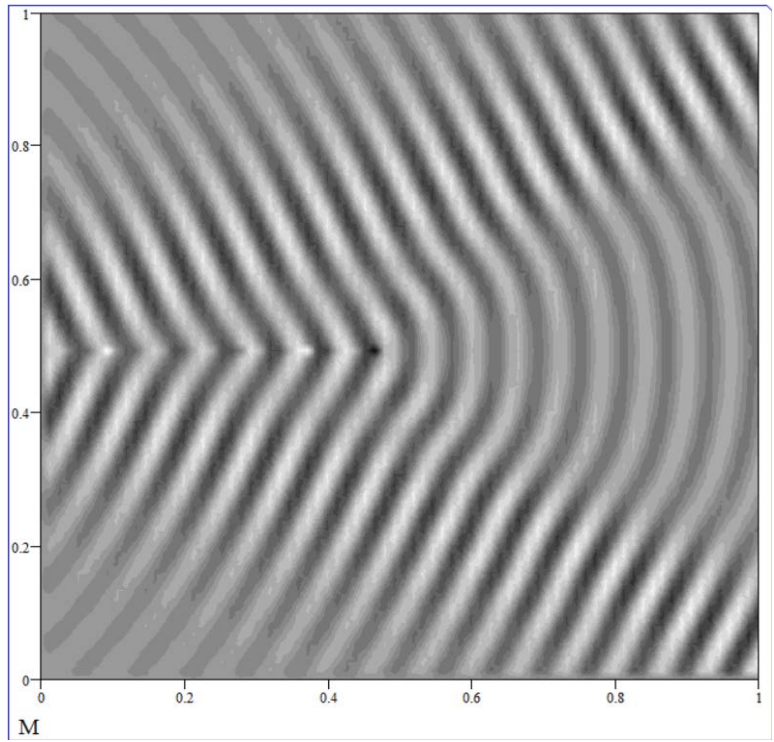

Figure 7. Spatial distribution of ground vibrations (in arbitrary units, greyscale) generated over the area of 75×75 m by a single load travelling on a straight track with acceleration $a = 0.5\ m/s^2$ and initial train speed $v_0 = 50\ m/s$.

It is seen from Figure 7 that on both sides of the track the concave wave fronts of Rayleigh waves are formed symmetrically, and the focusing occurs, accompanied by amplification of Rayleigh wave amplitudes in the focusing areas. Like in the case of curved tracks, this amplification is rather moderate, about two times. Note that the focusing areas (caustics) are moving away from the track with the distance, in agreement with the geometrical acoustics consideration discussed above.

Figure 8 shows the results of similar calculations of the spatial distribution of ground vibrations generated over the area of 75×75 m by a single axle load travelling along the straight track with a larger acceleration $a = 1\ m/s^2$ (other parameters are the same as in Figure 7). One can see that in this case of larger acceleration the focusing becomes more pronounced, as expected.

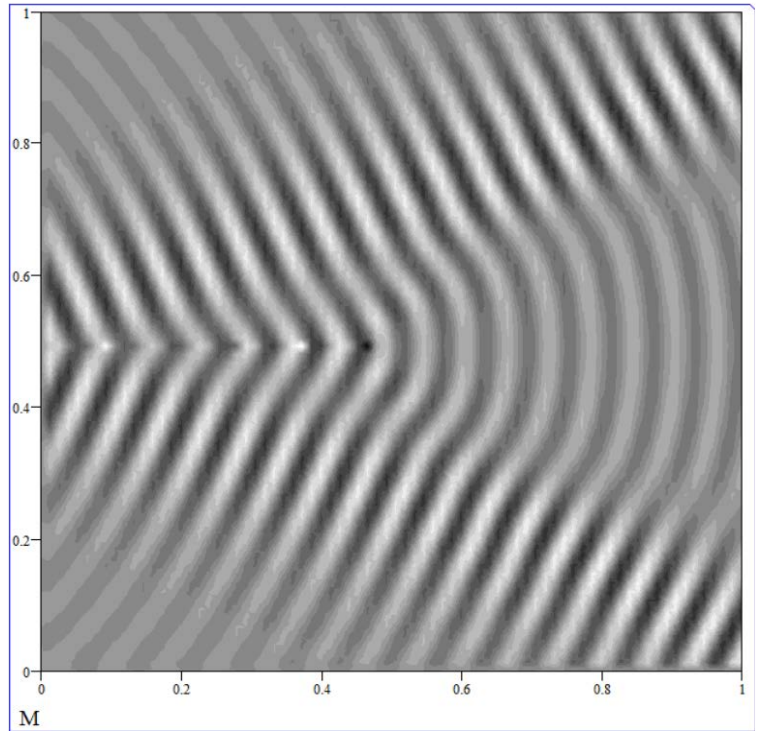

Figure 8. Spatial distribution of ground vibrations (in arbitrary units) generated over the area of 75×75 m by a single load travelling on a straight track with acceleration $a = 1\ m/s^2$ and initial train speed $v_0 = 50\ m/s$.

Figures 9 and 10 show the results of the calculations of the spatial distribution of ground vibrations generated over the area of 75×75 m by a single axle load travelling along the straight track with accelerations $a = 0.5\ m/s^2$ and $a = 1\ m/s^2$ respectively in the case of the lover value of the initial train speed $v_0 = 47\ m/s$ (other parameters are the same as in Figures 7 and 8). One can see that in this case of smaller initial velocity the focusing occurs closer to the track, as expected.

Note that focusing of railway-generated Rayleigh waves due to train acceleration affects both sides of a straight track. This effect should be taken into consideration by railway engineers and city planners when planning construction of new buildings in the areas close to the tracks where high speed trains move with acceleration and the current train speed is higher than Rayleigh wave velocity in the ground.

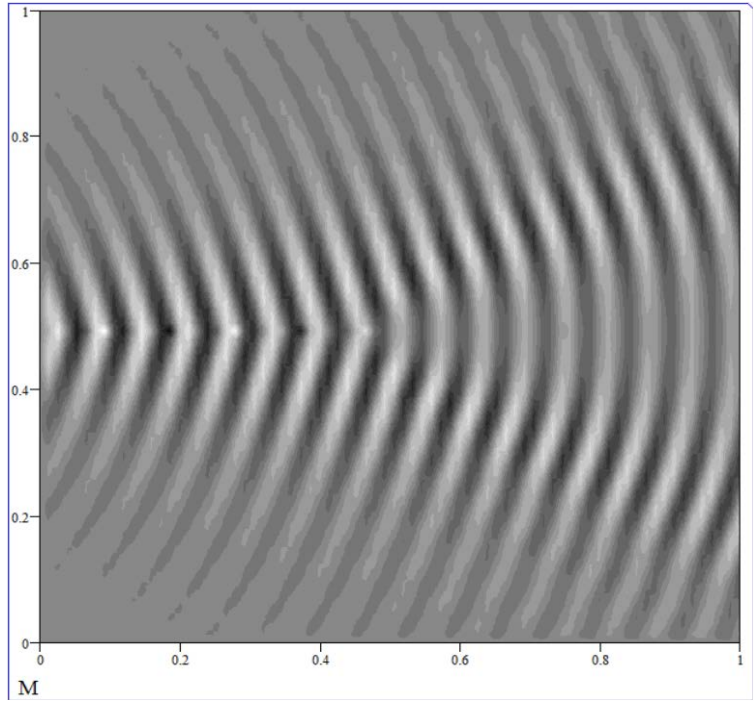

Figure 9. Spatial distribution of ground vibrations (in arbitrary units, greyscale) generated over the area of 75×75 m by a single load travelling on a straight track with acceleration $a = 0.5\ m/s^2$ and initial train speed $v_0 = 47\ m/s$.

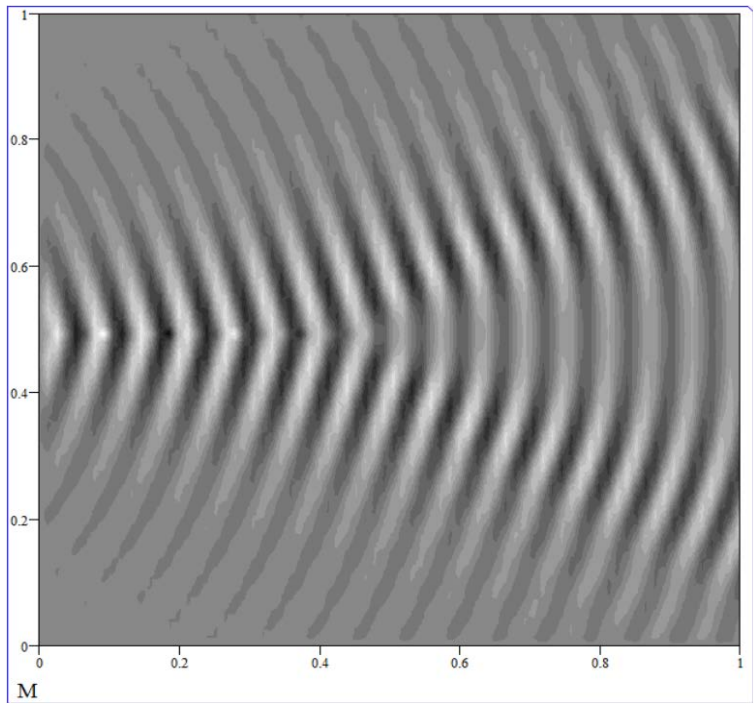

Figure 10. Spatial distribution of ground vibrations (in arbitrary units, greyscale) generated over the area of 75×75 m by a single load travelling on a straight track with acceleration $a = 1\ m/s^2$ and initial train speed $v_0 = 47\ m/s$.

# 5  Conclusions

It has been demonstrated in this paper that, for high-speed trains travelling along curved tracks at constant speeds under the condition of ground vibration boom (i.e. when the train speeds are larger than Rayleigh wave velocity in the ground), the effect of track curvature may result in focusing of railway-generated ground vibrations (Rayleigh surface waves) at one side of the track and in the corresponding increase of their amplitudes.

The focusing effect for Rayleigh waves generated by high-speed trains may occur also in the case of a train accelerating along a straight line if its current speed is larger than Rayleigh wave velocity in the ground. In this case the focusing occurs symmetrically on both sides of the track.

Railway engineers and city planners should be aware of the above-mentioned phenomena of focusing of Rayleigh ground waves generated by high-speed trains under the condition of ground vibration boom.